\documentclass[11pt,a4paper]{article}
\pdfoutput=1

\usepackage{jcappub}
\usepackage{bm}
\usepackage{comment}

\newcommand{\mg}{m_{3/2}}
\newcommand{\beq}{\begin{equation}}
\newcommand{\beqa}{\begin{eqnarray}}
\newcommand{\eeq}{\end{equation}}
\newcommand{\eeqa}{\end{eqnarray}}

\newcommand{\simgt}{\lower.5ex\hbox{$\; \buildrel > \over \sim \;$}}
\newcommand{\simlt}{\lower.5ex\hbox{$\; \buildrel < \over \sim \;$}}

\newcommand{\bd}[1]{\mbox{\boldmath $#1$}}

\begin{document}
\title{Cosmological Constraint on the Light Gravitino Mass from
CMB Lensing and Cosmic Shear}

\author[a,g]{Ken Osato,}
\author[b,c]{Toyokazu Sekiguchi,}
\author[d,g]{Masato Shirasaki,}
\author[e]{Ayuki Kamada,}
\author[a,f,g]{and Naoki Yoshida}

\affiliation[a]{Department of Physics, School of Science, The University of Tokyo, 7-3-1 Hongo, Bunkyo, Tokyo, 113-0033, Japan}
\affiliation[b]{University of Helsinki and Helsinki Institute of Physics, P.O. Box 64, FI-00014, Helsinki, Finland}
\affiliation[c]{Institute for Basic Science, Center for Theoretical Physics of the Universe, Daejeon 34051, South Korea}
\affiliation[d]{National Astronomical Observatory of Japan, Mitaka, Tokyo, 181-8588, Japan}
\affiliation[e]{Department of Physics and Astronomy, University of California, Riverside, California 92521, USA}
\affiliation[f]{Kavli Institute for the Physics and Mathematics of the Universe (WPI), 
U-Tokyo Institutes for Advanced Study, The University of Tokyo,
Kashiwa, Chiba, 277-8583, Japan}
\affiliation[g]{CREST, Japan Science and Technology Agency, 4-1-8 Honcho, Kawaguchi, Saitama, 332-0012, Japan}

\emailAdd{ken.osato@utap.phys.s.u-tokyo.ac.jp}

\abstract{
Light gravitinos of mass $\lesssim \mathcal{O} (10)$~eV are of particular interest in cosmology,
offering various baryogenesis scenarios without suffering from the
cosmological gravitino problem. The gravitino may contribute
considerably to the total matter content of the Universe
and affect structure formation from early to present epochs.
After the gravitinos decouple from
other particles in the early Universe, they free-stream 
and consequently
suppress density fluctuations of (sub-)galactic length scales.
Observations of structure at the relevant length-scales can be used to
infer or constrain the mass and the abundance of light gravitinos.
We derive constraints on the light gravitino mass
using the data of cosmic microwave background (CMB) lensing from Planck 
and of cosmic shear from the Canada France Hawaii Lensing Survey,
combined with analyses of the primary CMB anisotropies and
the signature of baryon acoustic oscillations in galaxy distributions.
The obtained constraint on the gravitino mass is
$\mg < 4.7$\,eV (95 \% C.L.), 
which is substantially tighter than the previous constraint from clustering analysis of
Ly-$\alpha$ forests.
}

\arxivnumber{1601.07386}

\maketitle
\flushbottom

\section{Introduction}
\label{sec:intro}
The existence of gravitino is predicted in particle physics models
with local supersymmetry (SUSY), or in supergravity models.
The gravitino mass $m_{3/2}$ is related to the energy scale of SUSY breaking, 
which is one of the most important quantities in low-scale SUSY models that are invoked
to solve the gauge hierarchy problem
(see, e.g., \cite{Martin:1997ns}, for a review).
However, it is well-known that gravitino can cause serious problems
in the cosmological context (the so-called cosmological gravitino problems~\cite{1982PhRvL..48.1303W}). 
If gravitino is stable (or very long-lived), 
its relic abundance may contradict with the observed energy density of dark matter 
unless $m_{3/2}$ is very small.
Gravitino with $m_{3/2}\lesssim \mathcal O(10)$\,eV 
has attracted particular attention, because such light gravitinos
can evade the cosmological gravitino problem~\cite{1993PhLB..303..289M}
and provide baryogenesis mechanisms
(e.g., thermal leptogenesis~\cite{1986PhLB..174...45F})
that work only at very high temperature.

Not only probed by collider experiments (e.g., LHC) with direct and indirect signatures,
the existence or the abundance of light gravitinos
has also been constrained from cosmological observations~\cite{1998PhRvD..57.2089P}.
Light gravitinos produced from thermal plasma in the early Universe
behave effectively as warm dark matter.
Gravitinos with sizable velocity dispersions
free-stream over a cosmological distance
until they become non-relativistic. 
Gravitinos suppress the growth of matter density fluctuations
and imprint characteristic signatures in the
matter power spectrum at and below the free-streaming length.
One can therefore constrain, in principle, the mass of light gravitinos 
from cosmological observations of large-scale matter distribution.
To present, observations of Ly-$\alpha$ forests~\cite{Viel2005} give
a stringent constraint 
on the gravitino mass as $m_{3/2} < 16$\,eV (95\% C.L.).
It is important to notice that cosmological constraints 
provide upper limits to $m_{3/2} < 16$\,eV, whereas
particle collider experiments generally
provide lower limits.

The objective of the present paper is to improve the cosmological constraint on $m_{3/2}$
using independent observations of CMB and the large-scale structure.
Weak gravitational lensing is a powerful probe of matter distribution.
The coherent distortion of images of distant galaxies can be used to infer the
foreground matter distribution. 
There are two background sources for weak gravitational lensing;
One is lensing of cosmic microwave background radiation, 
and the other is lensing of distant galaxies.
The difference in the source redshifts enables us to probe 
a wide spatial range of large-scale structure. 
Previous studies~\cite{Ichikawa2009,Kamada2014} suggest that
the two observables can be indeed utilized to constrain $m_{3/2}$.

There is another motivation to consider light gravitinos as a possible
matter content of the Universe.
It has been claimed that 
$\sigma_8$, the amplitude of matter density fluctuations normalized at $8\ h^{-1}\mathrm{Mpc}$,
derived from observations of CMB anisotropies
is higher than the values derived from observations of
large-scale structure at low redshifts, such as 
weak gravitational lensing and the abundance of galaxy clusters
\cite{2013JCAP...10..044H,2014PhRvL.112e1303B}.
The apparent tension may infer some mechanism that suppress
matter density fluctuations at late epochs. 
For example, massive active or sterile neutrinos~\cite{2013JCAP...10..044H,2014PhRvL.112e1303B,
2014MNRAS.444.3501B,Battye2015,MacCrann2015}, decaying dark matter~\cite{Enqvist2015},
and various baryonic physics~\cite{MacCrann2015,Osato2015} have been proposed.
The existence of light gravitinos may offer another intriguing
resolution for this tension, in quite a similar way to massive neutrinos.
Clearly, it is important to perform a fully consistent cosmological parameter estimate
including $m_{3/2}$ as one of the primary parameters.
This problem will be addressed in detail in appendix \ref{sec:sigma8}.
However, it is difficult to resolve this tension only with light gravitinos.

The rest of the paper is organized as follows. In section \ref{sec:theory}, we briefly summarize
particle physics aspects of the gravitino. We discuss how light gravitinos affect
large-scale structure in the Universe in section \ref{sec:lss}.
 In section \ref{sec:probes}, we describe the basics of two observational probes
that are used 
to derive constraints on gravitino mass $m_{3/2}$.
We explain in detail the observational data 
and methods to extract posterior distributions of cosmological parameters including
$m_{3/2}$ in section \ref{sec:data_method}.
In section \ref{sec:result}, we present constraints on the mass of light gravitinos
from these two observational probes combined with CMB and 
baryon acoustic oscillation (BAO) observations.
We give concluding remarks in section \ref{sec:conc}.

\section{Light gravitino}
\label{sec:theory}
Light gravitinos are realized typically in gauge-mediated SUSY
breaking (GMSB) scenarios~\cite{Dine:1981gu, Nappi:1982hm, AlvarezGaume:1981wy, Dine:1993yw, Dine:1994vc, Dine:1995ag}.
Let us review briefly basics of GMSB models and the current
constraints from LHC.
In GMSB models, a characteristic relation is supposed to hold among masses of the gravitino 
and SUSY particles (sparticles) in the Standard Model (SM) sector.
In fact, the discovery potential of GMSB models by LHC experiments
largely relies on these sparticle masses, 
and thus the current constraints on the gravitino mass are often model-dependent. 

The masses of the gravitino and sparticles originate from spontaneous SUSY breaking.
As a consequence, the masses are proportional to the SUSY breaking scale $\langle F \rangle$.
The gravitino mass is given by
\beq
m_{3/2} = \frac{| \langle F \rangle |}{\sqrt{3} M_{\rm pl}} \,,
\eeq
with $M_{\rm pl} \simeq 2.43 \times 10^{18}$\,GeV being the reduced Planck mass.
Sparticles in the SM sector acquire masses from the SUSY breaking through 
messenger fields, whose mass scale we denote as $M_{\rm mess}$.
With the gauge couplings denoted by
$g_{1}=\sqrt{5/3}g'$, $g_{2}=g$, and $g_{3}$, where $g'$ and $g$ are the conventional electro-weak 
gauge couplings, 
gaugino mass and sfermion mass squared are approximately given by
\beq
M_{a} \sim \frac{g_{a}^2}{16\pi^2} \frac{| \langle F \rangle |}{M_{\rm mess}} \quad \text{and} \quad 
m_{\tilde f_i}^2 \sim \sum_{a} C_{a}^{(i)} \left( \frac{g_{a}^2}{16\pi^2} \frac{| \langle F \rangle |}{M_{\rm mess}} \right)^2 \,,
\eeq 
respectively, where the indices $a$ and $i$ respectively indicate a SM gauge group
and a flavor of the sfermion, 
and $C_{a}^{(i)}$ is the quadratic Casimir invariant.
Lower bounds on their masses are placed by collider experiments
directly and indirectly as discussed below.
With some model-dependence, the mass constraints
can be translated into lower bounds on the SUSY breaking scale 
$| \langle F \rangle |$, or the gravitino mass $m_{3/2}$. 

Stringent constraints on GMSB models come from the Higgs mass $m_{h}=125$\,GeV 
measured by LHC~\cite{Aad:2015zhl,Chatrchyan:2012xdj}.
In the minimal supersymmetric Standard Model (MSSM),
the Higgs mass is bounded from above at tree level, $m_{h,{\rm tree}} < m_{Z}=91$\,GeV.
This requires a large contribution of radiative corrections from top-stop loops,
$m^{2}_{h,{\rm loop}} \sim m_{t}^4 / (16\pi^2 v^2) \ln\left(m_{\tilde t}/m_{t} \right)$,
where $m_{t}$ ($m_{\tilde t}$) and $v$ are the mass of top quark (stop) and
the vacuum expectation value of Higgs, respectively.
To achieve the observed large Higgs mass,
the stop mass is required to be as large as $m_{\tilde t}=\mathcal O(10\text{--}100)$\,TeV, which
correspondingly places a lower bound on the gravitino mass.
For example, in a class of GMSB models with $N_5$ copies of messenger fields
in the ${\bf 5}+{\overline {\bf 5}}$ representation of $SU(5)$,\footnote{
In order to maintain perturbative unification of the SM gauge couplings, $N_{5}$ 
needs to be not too large (typically $N_{5} \leq 5$). For details of the model, especially 
the explicit formulas of sparticle masses, we refer to, e.g., \cite{Martin:1997ns,Kamada2014}.
}
one obtains a bound $m_{3/2}>300$\,eV with $N_{5}=1$ ($60$\,eV with $N_5=5$)~\cite{Ajaib:2012vc}, provided
that the coupling between the messengers and a SUSY breaking field $\lambda$ is perturbative (i.e., $|\lambda|<1$).
As will be discussed in the next section, such light gravitinos are ruled out or only marginally allowed 
in order for their thermal relic density not to exceed the observed density of dark matter.
However, the bound is model-dependent, and $m_{3/2}=\mathcal O(1\text{--}10)\,{\rm eV}$ may be possible
if the coupling is non-perturbative $|\lambda|>1$ or a singlet Higgs (i.e., next to MSSM) is introduced~\cite{Yanagida:2012ef}.
Interestingly, the former may offer a hidden baryon as 
a main component of dark matter~\cite{1996PhLB..389...37D}.

Less stringent lower bounds on the gravitino mass are obtained also from direct SUSY searches in
LHC that seek for production of sparticles (mostly squarks and gluinos) decaying
into energetic SM particles and gravitinos. For example, in the same GMSB model
as mentioned above, 
with $M_{\rm mess}=250$\,TeV and $N_{5}=3$ (${\bf 10}+{\overline {\bf 10}}$ of $SU(5)$) being fixed, 
a bound $|\lambda \langle F \rangle| / M_{\rm mess}>63$\,TeV is obtained 
from analysis of events with at least one tau lepton and zero or one light lepton in $20\,{\rm fb}^{-1}$ 
of the LHC $8$\,TeV run~\cite{Aad:2014mra}. Assuming a perturbative coupling 
$|\lambda|<1$, this bound leads to $m_{3/2} > 3.7$\,eV. We also note that in the future 
International Linear Collider may allow us to measure the light gravitino mass directly
from decay of a ``long-lived" next-to-lightest supersymmetric particle~\cite{Matsumoto:2011fk,Katayama:2013}.

\section{Effects on large-scale structure of the Universe}
\label{sec:lss}
In what follows, we assume that the reheating temperature is so high that light gravitinos are once
in equilibrium with a thermal bath in the very early universe.
As the background temperature decreases, the gravitino decouples from other particles
at some point,
and its relic abundance in the Universe is fixed.
The relic abundance is approximately estimated from the relativistic degrees
of freedom (hereafter denoted as $g_{*s3/2}$) in 
a thermal plasma at the gravitino decoupling. 
In~\cite{1998PhRvD..57.2089P,Ichikawa2009}, 
the Boltzmann equation is solved for the gravitino number density in GMSB models.
For a messenger mass scale $M_{\rm mess}\sim100$\,TeV, 
$g_{*s3/2} \sim 90$ with only mild dependence on $m_{3/2}$ in
$1\text{--}100$\,eV~\cite{Ichikawa2009}.
Throughout the present paper, we set $g_{*s3/2}=90$ as a canonical value.

At late times, thermal relic gravitinos act as warm dark matter particles
that can be characterized by the temperature (velocity dispersion) and mass.
The phase-space distribution of the gravitinos
is given by the Fermi-Dirac distribution with two degrees of freedom because
only the (spin $\pm$1/2) goldstino components virtually interact with other particles.
If there is no significant entropy production after the gravitino decoupling, 
the gravitino temperature is given in terms of the standard neutrino
temperature $T_\nu=1.95$\,K as
\beq
T_{3/2}=\left(\frac{g_{*s\nu}}{g_{*s3/2}}\right)^{1/3}T_\nu=0.96~\mbox{K}\left(\frac{g_{*s3/2}}{90}\right)^{-1/3},
\eeq
where $g_{*s\nu}=10.75$ is the relativistic degree of freedom at the neutrino decoupling. 
Then the effective number of neutrino species accounting for the gravitino is given by
\beq
N_{3/2} =
\left( \frac{T_{3/2}}{T_\nu} \right)^4
= \left( \frac{g_{*s\nu}}{g_{*s3/2}} \right)^{4/3}
=0.059\left(\frac{g_{*s3/2}}{90}\right)^{-4/3}.
\eeq
At late epochs, light gravitinos are non-relativistic, so that
the energy density can be estimated as
\beq
\Omega_{3/2} h^2 = 0.13 \left( \frac{m_{3/2}}{100\ \mathrm{eV}} \right) \left( \frac{g_{*s3/2}}{90} \right)^{-1}.
\eeq

We note that dark matter cannot consists solely of the gravitino
 in the cosmological model with thermally produced gravitinos considered here.
This is because $m_{3/2}$ needs to be as large as 86\,eV in order to account for the observed 
dark matter density $\Omega_{3/2} h^2 \simeq0.11$~\cite{PlanckParameters}, which clearly contradicts 
the existing constraint from Ly-$\alpha$ forest, $m_{3/2}<16$\,eV~\cite{Viel2005}.
In what follows, we assume that dark matter consists of the light gravitino
and some additional CDM constituents, so that
\beq
\Omega_\mathrm{dm} = \Omega_\mathrm{cdm} + \Omega_{3/2}.
\eeq
For example, within the framework of GMSB, where the gravitino is usually the lightest SUSY particle, 
a messenger baryon or QCD axion can be the extra CDM.

The primary effects of light gravitinos on structure formation
are of twofold.
The epoch of matter-radiation equality is slightly delayed (corresponding to a larger $a_{\rm eq}$),
and the matter fluctuations at small length scales are suppressed.
Regarding the first effect, light gravitinos are relativistic in the early universe
and can contribute to the radiation energy.
However, the contribution of the light gravitinos is too small ($N_{3/2}\simeq 0.059$)
to change the epoch of the equality appreciably when compared to current observational
sensitivities\footnote{
For reference, the current constraint on the effective number of massless neutrino species is 
$N_{\rm eff}=3.04\pm 0.2$ from CMB and baryon acoustic oscillation (BAO) 
in galaxy distributions~\cite{PlanckParameters}, which is not sensitive enough 
to measure $\Delta N_{\rm eff}=0.059(=N_{3/2})$.
Furthermore, if the neutrino species has a total mass of $\mathcal{O} (1)$\,eV,
the constraint on $\Delta N_{\rm eff}$ should be even less stringent since such species
are already more or less non-relativistic at the equality.
}.

In order to constrain the mass of light gravitino, we consider the latter effect.
Light gravitinos free-stream with a sizable velocity, in a similar manner to massive neutrinos. 
Within the free-streaming scale, gravitinos do not cluster,
and thus act effectively as ``drag'' of the growth of matter fluctuations.
Thus, typically, matter fluctuations at late times are smaller
than in the conventional CDM model.
The characteristic length scale can be estimated as~\cite{Kamada2014}
\beq
\label{eq:fs}
k_\mathrm{J} = \left. a \sqrt{\frac{4\pi G \rho_\mathrm{m}}{\langle v^2 \rangle}} \right|_{a=a_\mathrm{eq}}
\simeq 0.86 \ \mathrm{Mpc}^{-1}
\left( \frac{m_{3/2}}{100\ \mathrm{eV}} \right) ^{1/2}
\left( \frac{g_{*s3/2}}{90} \right)^{5/6}.
\eeq
Massive neutrinos affect the growth of structure, but
the temperature and the energy density of massive neutrinos
are different from those of the light gravitinos. For massive neutrinos
the resulting suppression scale eq. \ref{eq:fs} differs.

\begin{figure}
\centering \includegraphics[clip, width=0.8\textwidth]{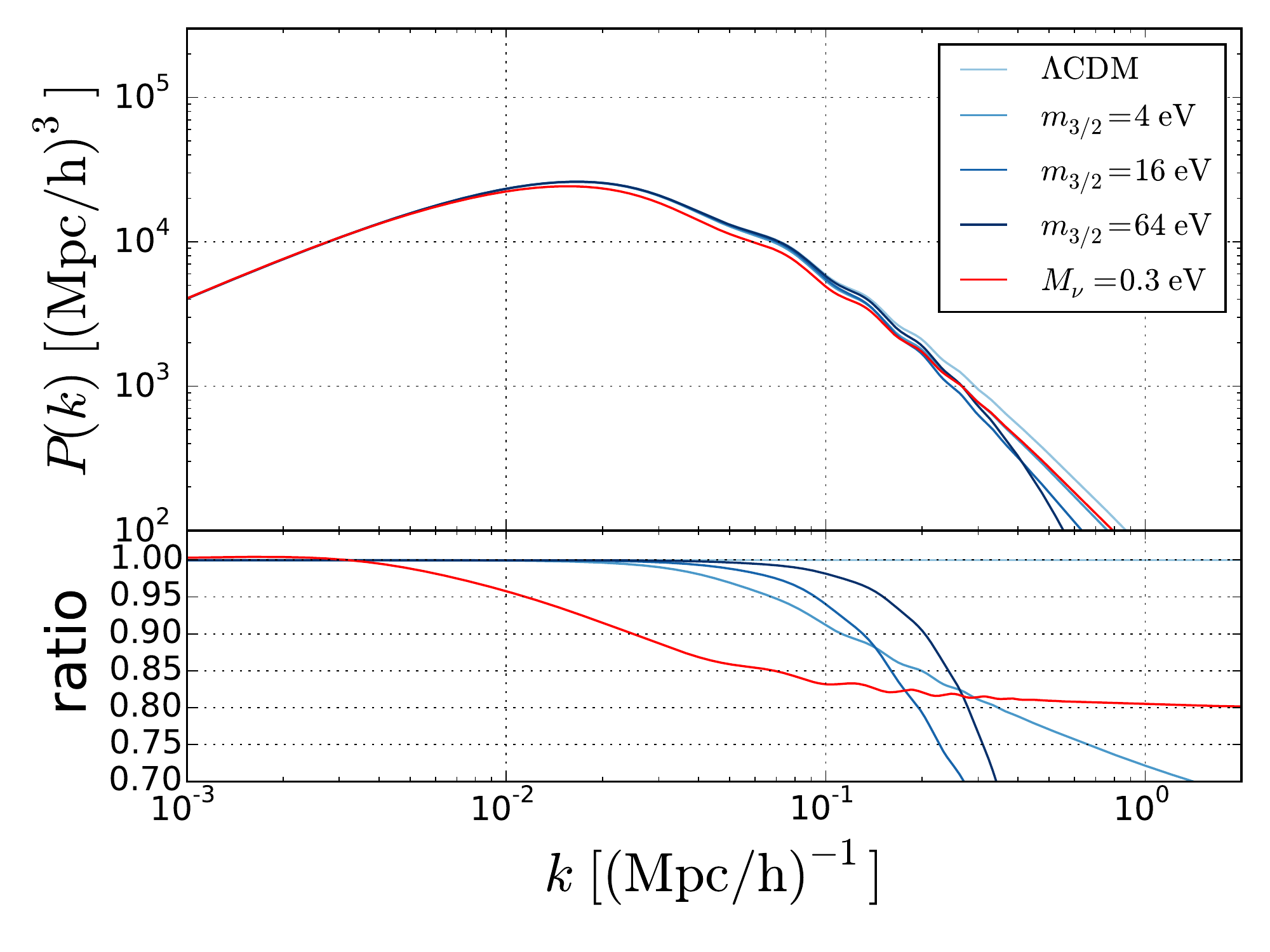}
\caption{
  Linear matter power spectra at $z=0$ for three different values of $m_{3/2}$
  computed with the {\tt CLASS} code.
  For comparison, we also show the spectrum with massive neutrinos
  with the total mass $M_\nu = 0.3\,\mathrm{eV}$.
  The total dark matter density $\Omega_\mathrm{dm} = \Omega_\mathrm{cdm} + \Omega_{3/2}$
  or  $\Omega_\mathrm{dm} = \Omega_\mathrm{cdm} + \Omega_{\nu}$
  is fixed to be $\Omega_\mathrm{cdm}$ of
  the result of Planck 2015 TT,TE,EE+lowP dataset~\cite{PlanckParameters}.
  Upper panel shows the absolute values and lower panel shows the fractional differences 
  with respect to the standard $\Lambda$CDM.
  }
\label{fig:grav_pk}
\end{figure}

To demonstrate the effect of the free-streaming of light gravitinos,
in figure~\ref{fig:grav_pk} we plot
linear matter power spectra at $z=0$ in the presence of the light gravitino with different masses.
The power spectra are computed using the Boltzmann code {\tt CLASS}~\cite{Lesgourgues2011,Blas2011}
with the base cosmological parameters of
the Planck 2015 TT,TE,EE+lowP dataset~\cite{PlanckParameters}.
At large scales, the linear matter power spectra are 
rather insensitive to $m_{3/2}$. 
The characteristic suppression wavenumber
is larger for larger $m_{3/2}$, because 
the light gravitinos become non-relativistic earlier
and the free-streaming length becomes smaller.
On the other hand, the fraction of gravitinos in the matter content
increases with increasing $m_{3/2}$,
and then the power is more effectively suppressed. 
The figure suggests that, in order to probe the gravitino
mass of our interest (i.e., $m_{3/2}=\mathcal O(1\text{--}10)$\,eV),
we need a cosmological probe that is sensitive
to matter power spectrum around $k=\mathcal O(0.01\text{--}0.1)$\,Mpc$^{-1}$.

\section{Observational probes}
\label{sec:probes}
Weak gravitational lensing effect is one of the most powerful 
probes of matter power spectrum at scales $k=\mathcal O(0.01\text{--}0.1)$\,Mpc$^{-1}$.
To constrain the gravitino mass, we use recent data from
CMB lensing and cosmic shear observations. 
In the following, we briefly review these observations in order.
Following the previous section, we in this section adopt the cosmological parameters from the 
Planck 2015 TT,TE,EE+lowP dataset~\cite{PlanckParameters}
with fixed $\Omega_\mathrm{dm}=\Omega_{\rm cdm}+\Omega_{3/2}$.

\begin{description}
\item {\bf CMB lensing}
CMB photons travel through the gravitational potential
between the last scattering surface and the observer point.
The collective gravitational lensing effect
is called CMB lensing and, in principle, 
contains information of the matter distribution in the Universe at 
$z\simeq2-3$ (e.g., see ref.~\cite{Lewis2006} for a review).
The CMB lensing is described by the lensing potential $\phi (\hat{n})$,
\beq
\phi (\hat{n}) = -2 \int_0^{\chi_*} {\rm d}\chi \ \frac{f_K(\chi_* - \chi)}{f_K(\chi_*) f_K(\chi)} \Psi (\chi \hat{n},\eta(\chi)), \label{eq:def_lenspot}
\eeq
where $\chi$ is a radial comoving distance from us,
$\chi_*$ is a comoving distance to the last scattering surface, 
$\eta (\chi)$ is a conformal time
at which the photon is at $\chi \hat{n}$, 
$\Psi$ is a gravitational potential.
$f_K (\chi)$ is a comoving angular diameter distance,
\beq
f_K(\chi) =
\begin{cases}
  K^{-1/2} \sin (K^{1/2}\chi) & (K > 0) \\
  \chi & (K = 0) \\
  (-K)^{-1/2} \sin [(-K)^{1/2}\chi] & (K < 0),
\end{cases}
\eeq
where $K$ is a curvature of the Universe.
Throughout this paper, we assume a flat universe with $K=0$.

The power spectra of the lensing potential contains rich information of
the matter distribution at intermediate redshifts.
The lensing potential can be expanded with spherical harmonics as
\beq
\phi (\hat{n}) = \sum_{\ell ,m} \phi_{\ell m}Y_{\ell m}(\hat{n}).
\eeq
The angular power spectrum of $\phi$ is then defined by
\beq
\langle \phi_{\ell m} \phi_{\ell' m'} \rangle = \delta_{\ell \ell'} \delta_{m m'} C_\ell^{\phi \phi}. \label{eq:def_Cphiphi}
\eeq
We can calculate $C_\ell^{\phi \phi}$ by using eq.~\ref{eq:def_lenspot}
and Limber approximation~\cite{Kaiser1992}
\footnote{The exact expression of $C_{\ell}^{\phi \phi}$ is found in \cite{Hu:2000ee}, while eq.~\ref{eq:model_Cphiphi} is an accurate approximation for $\ell \simgt 10$ where we are interested in.} as follows:
\beqa
C_{\ell}^{\phi \phi} = \int_0^{\chi_*} \frac{{\rm d}\chi}{f_K(\chi)^2} \, 
\left[-2\frac{f_K(\chi_* - \chi)}{f_K(\chi_*) f_K(\chi)}\right]^2 
P_{\Psi} (\ell/f_K(\chi), \eta(\chi)), \label{eq:model_Cphiphi}
\eeqa
where $P_{\Psi} (k, \eta(\chi))$ is the power spectrum of 
the gravitational potential, related with the matter density through
the Poisson equation.

In eq.~\ref{eq:model_Cphiphi}, it is sufficient to consider fluctuations
to linear order,
because we mainly focus on the range of $40 < \ell < 400$ where
non-linear gravitational growth is unimportant 
\cite{Hu:2000ee}.
With the linear approximation, $P_{\Psi}$
is expressed as
\beq
P_{\Psi} (k, \eta) = 
\left| T_{\Psi} (k, \eta)\right|^2 P_{\cal R}(k), \label{eq:Pow_psi_lin}
\eeq
where $T_{\Psi}$ is the transfer function and 
$P_{\cal R}$ represents the power spectrum of 
the primordial curvature fluctuations.

\item {\bf Cosmic shear}
Images of distant galaxies are 
distorted by weak gravitational lensing effect 
of foreground matter distributions.
One can characterize the distortion 
by the following 2D matrix:
\beqa
A_{ij} = \frac{\partial \beta_{i}}{\partial \theta_{j}}
           \equiv \left(
\begin{array}{cc}
1-\kappa -\gamma_{1} & -\gamma_{2}  \\
-\gamma_{2} & 1-\kappa+\gamma_{1} \\
\end{array}
\right), \label{distortion_tensor}
\eeqa
where $\kappa$ is convergence,
$\gamma$ is shear, and
$\bd{\theta}$ and $\bd{\beta}$ represent the observed position 
and the true position of a source, respectively.
In weak lensing limit (i.e., $|\kappa|, |\gamma| \ll 1$), 
each component of $A_{ij}$ can be related to
the second derivative of the gravitational potential $\Psi$ 
\cite{Bartelmann:1999yn,Munshi:2006fn}.

By using the Poisson equation, one can relate the convergence
field to the matter overdensity field $\delta$ as
\cite{Bartelmann:1999yn,Munshi:2006fn}
\beqa
\kappa(\hat{n})= 
\int _{0}^{\chi_H}{\rm d}\chi G(\chi)
\delta[\chi \hat{n}, \eta(\chi)], \label{eq:kappa_delta}
\eeqa
where $G(\chi)$ is the lensing efficiency.
Given a source galaxy distribution $p(\chi)$ normalized to $\int d\chi\,p(\chi)=1$, 
$G(\chi)$ is given by
\beqa
G(\chi) = \frac{3}{2} \frac{\Omega_\mathrm{m}}{a(\chi)} \left( \frac{H_0}{c} \right)^2 f_K(\chi)
\int_\chi^{\chi_H} {\rm d}\chi'\, p(\chi') \frac{f_K(\chi'-\chi)}{f_K(\chi')},
\eeqa
where $\chi_H$ is a comoving distance to the horizon.
The convergence power spectrum can then be computed as
\beq
C^{\kappa\kappa}_\ell = \int_0^{\chi_H} \frac{{\rm d}\chi}{f_K^2(\chi)} \, G^2(\chi) P_\delta \left(\ell/f_K (\chi), \eta(\chi) \right),
\eeq
where $P_\delta$ is the three dimensional matter power spectrum.
The direct observables are the two-point correlation functions (2PCFs)
of cosmic shear, $\xi_\pm (\vartheta)$ (see ref.~\cite{Schneider:2002jd} for details).
Via Hankel transformation, $\xi_\pm (\vartheta)$ can be related with $C^{\kappa\kappa}_\ell$ as
\beq
\xi_{+,-} (\vartheta) = \int_0^\infty \frac{\ell d\ell}{2\pi} J_{0,4}(\ell \vartheta) 
C^{\kappa\kappa}_\ell,
\eeq
where the plus (minus) sign corresponds to the Bessel function of the first kind $J_0$ ($J_4$).
\end{description}

We calculate convergence power spectra and CMB lensing potential spectra
using {\tt CLASS}.
Non-linear correction is computed by the {\tt HALOFIT} fitting
formula~\cite{Takahashi2012}
modified to incorporate massive neutrinos~\cite{Bird2012}.
At late epochs, both massive neutrinos and light gravitinos are non-relativistic and
behave similarly.
We can easily modify the code based on the one with massive neutrinos
to incorporate the effect of light gravitinos.
We simplify call the modified one {\tt HALOFIT} in the following.

\begin{figure}
\begin{center}
\includegraphics[clip, width=0.8\textwidth]{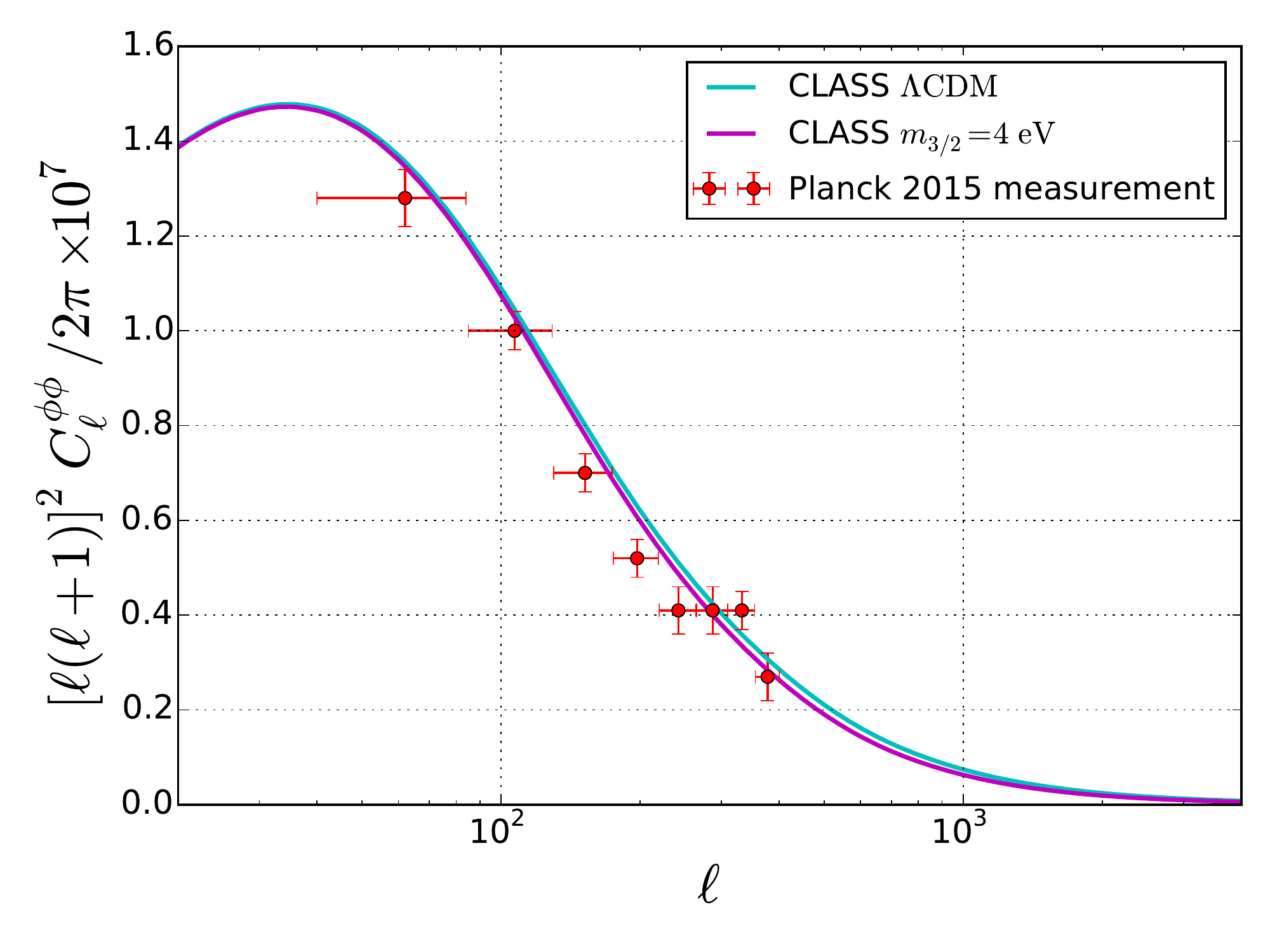}
\end{center}
\caption{
	CMB lensing auto-power spectrum in the presence of light gravitinos.
	The gravitino mass is set to be 0\,eV (cyan line) and 4\,eV (magenta line) as in 
	figure~\ref{fig:grav_pk}.
	The red points show the Planck 2015 measurement
	\cite{PlanckLensing}.
	We adopt cosmological parameters from the Planck 2015 TT,TE,EE+lowP 
	dataset~\cite{PlanckParameters} for this plot.
	The offset of the data points and the model curves, which adopt Planck 2015 TT,TE,EE+lowP parameters,
	can be interpreted as the often-claimed tension in $\sigma_{8}$ between
	observations of CMB anisotropies of temperature and polarizations
	and observations of large-scale structure.
}
\label{fig:grav_cmb}
\end{figure}

\begin{figure}
\begin{center}
\includegraphics[clip, width=1.0\textwidth]{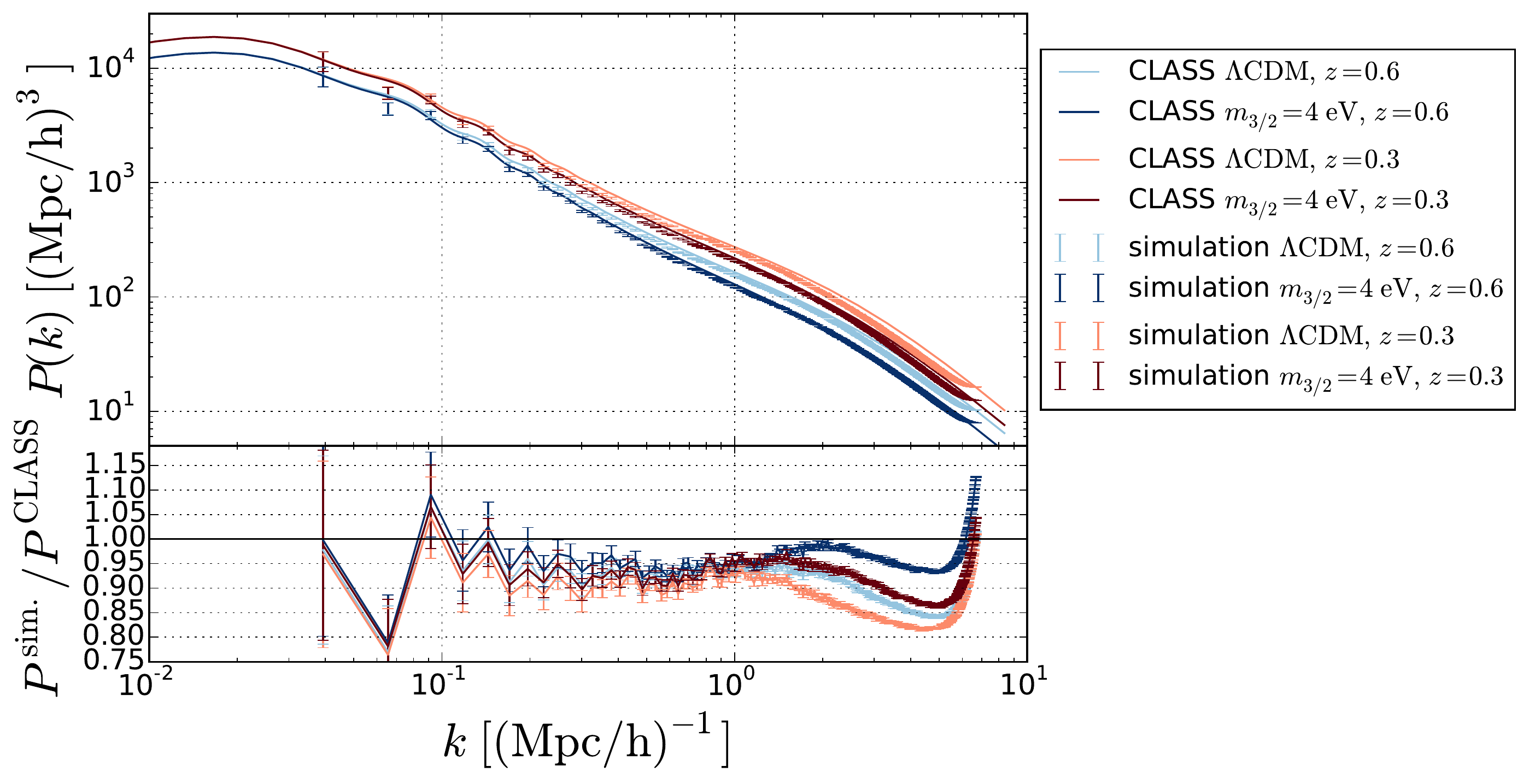}
\end{center}
\caption{
	We compare the non-linear matter power spectra calculated by
	{\tt CLASS} (solid lines) with {\tt HALOFIT} and the simulation results 
	of \cite{Kamada2014} (points with error bars) at $z=0.3,\ 0.6$.
	Only in this plot, we adopt cosmological parameters from Planck 2013 temperature-only results
	\cite{Planck2013Parameters} because $N$-body simulations in \cite{Kamada2014}
	adopted Planck 2013 parameters.
}
\label{fig:grav_pk_nl}
\end{figure}

\begin{figure}
\begin{center}
\includegraphics[clip, width=1.0\textwidth]{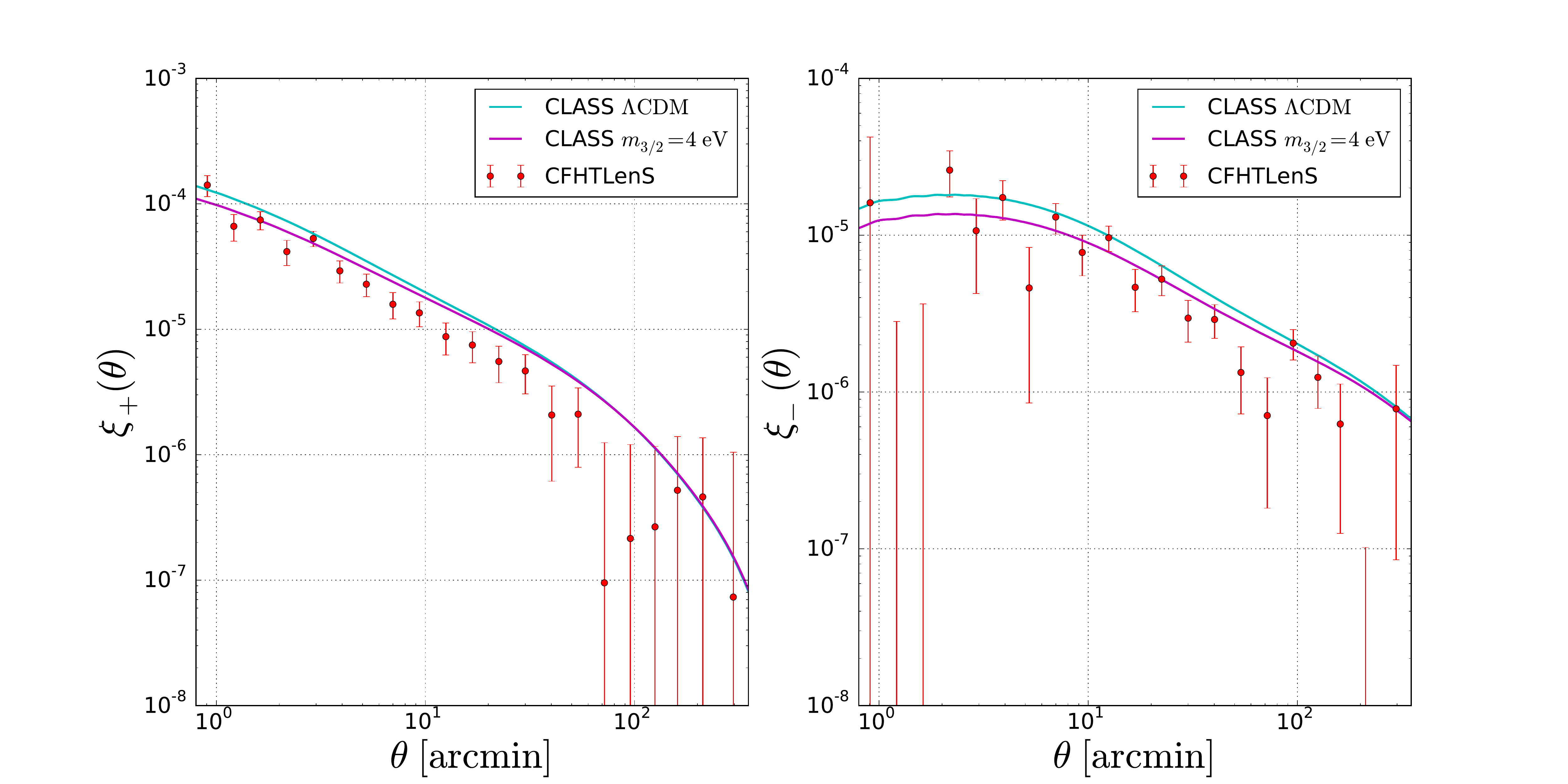}
\end{center}
\caption{
	We compare the 2PCFs of cosmic shear computed with {\tt CLASS} (solid lines)
	and the measurements of the CFHTLenS~\cite{Kilbinger2013} (red points).
	The left (right) panel shows $\xi_+ (\theta)$ ($\xi_- (\theta)$).
	The cyan (magenta) line shows the result with the gravitino mass 0\,eV (4\,eV).
	In the model calculations, the redshift distributions of the source galaxies
	are included as the weight function.
	We adopt the cosmological parameters from
	Planck 2015 TT,TE,EE+lowP dataset~\cite{PlanckParameters}.
	Notice the apparent tension in $\sigma_{8}$ as in figure~\ref{fig:grav_cmb}.
}
\label{fig:grav_xi}
\end{figure}

In figure~\ref{fig:grav_cmb}, we show the power spectra of CMB lensing
potential $C^{\phi\phi}_\ell$
computed with different masses of the light gravitinos.
We also plot the measurement of the Planck mission~\cite{PlanckLensing}.
Figure~\ref{fig:grav_pk_nl} compares the non-linear matter
power spectra calculated using {\tt CLASS} with {\tt HALOFIT}
and the $N$-body simulation results of \cite{Kamada2014} at $z=0.3,\ 0.6$
in order to check the validity of the {\tt HALOFIT} treatment.
These two results are consistent within 10\% level.
In figure~\ref{fig:grav_xi}, the comparison of 2PCFs of cosmic shear calculated from {\tt CLASS}
with measured 2PCFs by the CFHTLenS survey~\cite{Kilbinger2013} are shown.
We find suppression of the power spectrum and of the correlation function
for models with light gravitinos.

\section{Data set and parameter estimation}
\label{sec:data_method}
To estimate cosmological parameters including
the mass of light gravitinos, 
we perform the Markov chain Monte Carlo (MCMC) analysis.
For this purpose, we make use of the publicly available code
{\tt MontePython}~\cite{Audren2013} with the Metropolis-Hastings sampling.
The data used in this study are summarized below.

\subsection*{Planck TT,TE,EE+lowP}
The Planck satellite provides CMB anisotropy maps 
along with a subset of the polarization data.
The measurement is done with the 
angular resolution of $\sim10$ arcmin 
at the frequencies of 25--1000 GHz, 
allowing the estimation of CMB power spectra 
for $\ell\simlt2000$ without significant contaminants 
from astrophysical sources.
We use measurements of the angular power spectra of 
CMB temperature and polarizations anisotropies made by
Planck.
The dataset consists of the three auto power spectra 
of the temperature ($C_{\ell}^{\rm TT}$), 
the E-mode ($C_{\ell}^{\rm EE}$)
and the B-mode polarizations ($C_{\ell}^{\rm BB}$),
and the cross power spectrum of the temperature 
and the E-mode polarization ($C_{\ell}^{\rm TE}$).
The ranges of multipole are $\ell =2$--$2508$ for 
$C_{\ell}^{\rm TT}$, 
$\ell =2$--$1996$ for $C_{\ell}^{\rm TE}$ and 
$C_{\ell}^{\rm EE}$, and $\ell=2$--$29$ for $C_{\ell}^{\rm BB}$.
We use the publicly available Planck likelihood code~\cite{PlanckLikelihood}.

\subsection*{Planck lensing}
The CMB maps provided by the Planck satellite 
enable us to estimate the lensing potential over approximately 
70\% of the sky~\cite{PlanckLensing}.
The reconstruction of the lensing potential from the observed CMB maps
has been performed by the quadratic estimator~\cite{Okamoto:2003zw}.
Details of the reconstruction are described in \cite{PlanckLensing}.
The reconstructed lensing potential has been detected at a significance 
of $\sim40\sigma$ from the combined analysis 
with the CMB temperature and polarizations data.
We use the angular auto power spectrum of lensing potential 
($C_{\ell}^{\phi \phi}$) 
in the range of multipole of $\ell=40$--$400$.

\subsection*{CFHTLenS 2PCFs}
The Canada France Hawaii Lensing Survey (CFHTLenS) 
is an imaging survey in five optical bands with the sky coverage 
of 154 square degrees.
The CFHTLenS achieved an effective weighted number density 
of $\sim11$ galaxies per square arcminutes 
with shape and photometric redshift estimates~\cite{Erben:2012zw},
allowing robust and accurate weak lensing analysis~\cite{Heymans2012}.
We use the 2PCFs of galaxy's ellipiticities in the CFHTLenS measured
by \cite{Kilbinger2013}.
The measured 2PCFs are known as the good estimator of $\xi_\pm$
in the absence of intrinsic alignment (IA) of galaxies 
\cite{Schneider:2002jd}.
In \cite{Kilbinger2013}, 2PCFs have been measured for the source galaxies
with the redshift of $0.2<z<1.3$.
It is expected that the broad redshift distribution of source galaxies  
makes the IA contribution sub-dominant (see, e.g., \cite{Kirk:2011aw}).
We check that the effect of IA is negligible by removing the data
of small angular separation (see section \ref{sec:result}).
The correlation functions are binned with 21 bins
in the range $0.9\ \mathrm{arcmin} \leq \theta \leq 300\ \mathrm{arcmin}$.
We consider a Gaussian likelihood function of the 2PCFs with
the covariance matrix as in \cite{Kilbinger2013}.

\subsection*{BOSS}
The Baryon Oscillation Spectroscopic Survey (BOSS) is 
designed  to  obtain spectra and redshifts for 1.35 million 
galaxies covering 10,000 square degrees in 
the Sloan Digital Sky Survey (SDSS)~\cite{York:2000gk}.
BOSS includes the galaxy catalog 
from SDSS data releases 10~\cite{Ahn:2013gms}
and 11~\cite{Alam:2015mbd}, 
allowing us to select red galaxies 
in the spectroscopic redshift range of $0.2<z<0.7$.
The BAO feature has been detected in the clustering of galaxies from the BOSS
at a significance of $\simgt7\sigma$~\cite{Anderson2014},
and it has been 
fully utilized to measure the cosmic distance scale with a 1\% precision.
We use the measurement of the cosmic distance scale in \cite{Anderson2014}
to improve the cosmological constraints derived 
from the CFHTLenS.
We consider the ratio of the cosmic distance measure $D_V$
and the sound horizon at the drag epoch $r_s$. 
The measure of $D_V/r_s$ is performed for 
two galaxy samples named LOWZ ($z_\mathrm{eff} = 0.32$) 
and CMASS ($z_\mathrm{eff} = 0.57$).
The result is given by \cite{Anderson2014}
\beqa
D_V/r_s (\mathrm{LOWZ}) &=& 8.47 \pm 0.17, \\ 
D_V/r_s (\mathrm{CMASS}) &=& 13.77 \pm 0.13.
\eeqa

We explore two cosmological models with and without light gravitinos.
As a base model without gravitinos,
we assume the conventional flat power-law $\Lambda$CDM model with six parameters
($\Omega_\mathrm{cdm}h^2$, $\Omega_\mathrm{b}h^2$, $100\theta_s$, $\ln (10^{10}A_s)$, $n_s$, $\tau_\mathrm{reio}$).
As its extension, $\Lambda$CDM+gravitino model has an additional parameter $m_{3/2}$.
We assume that
all neutrinos are massless with the effective degree of freedom $N_\mathrm{eff} = 3.046$. 
Because massive neutrinos and light gravitinos affect the matter power spectrum
in a similar manner~\cite{Kamada2014}, and their effects are roughly additive,
the upper bounds on gravitino mass we derive in the following are conservative.
Theoretical nonlinear matter power spectrum is computed by {\tt CLASS} with
the {\tt HALOFIT} nonlinear corrections.

We note that there is one caveat associated with the applicability
of the {\tt HALOFIT} formula to our model.
Originally, it is calibrated by simulations with the total neutrino mass $M_\nu \lesssim 0.6$\,eV~\cite{Bird2012}. 
It is not clear if the same formula applies to models with light gravitinos
with mass $\gtrsim \mathcal O(1)$\,eV.
However, the results of {\tt HALOFIT} and N-body simulations with light gravitino of $m_{3/2}=4$\,eV
show reasonable agreement (figure~\ref{fig:grav_pk_nl})
at observable angular scales of $\ell \lesssim\mathcal O(100)$.\footnote{
Angular scales of CMB lensing measurements are much larger and nonlinear corrections are even less relevant.
} In addition, as we shall see in the next section,
light gravitinos with a significantly larger mass, $m_{3/2}\gtrsim 10$\,eV is
already disfavored observationally in light of the Planck (TT,TE,EE+lowP)+lensing.
Thus we expect that using the {\tt HALOFIT} model gives reasonably accurate result
in the parameter ranges we consider here.

\section{Constraints on gravitino mass}
\label{sec:result}

\begin{figure}
\centering \includegraphics[clip, width=1.0\textwidth]{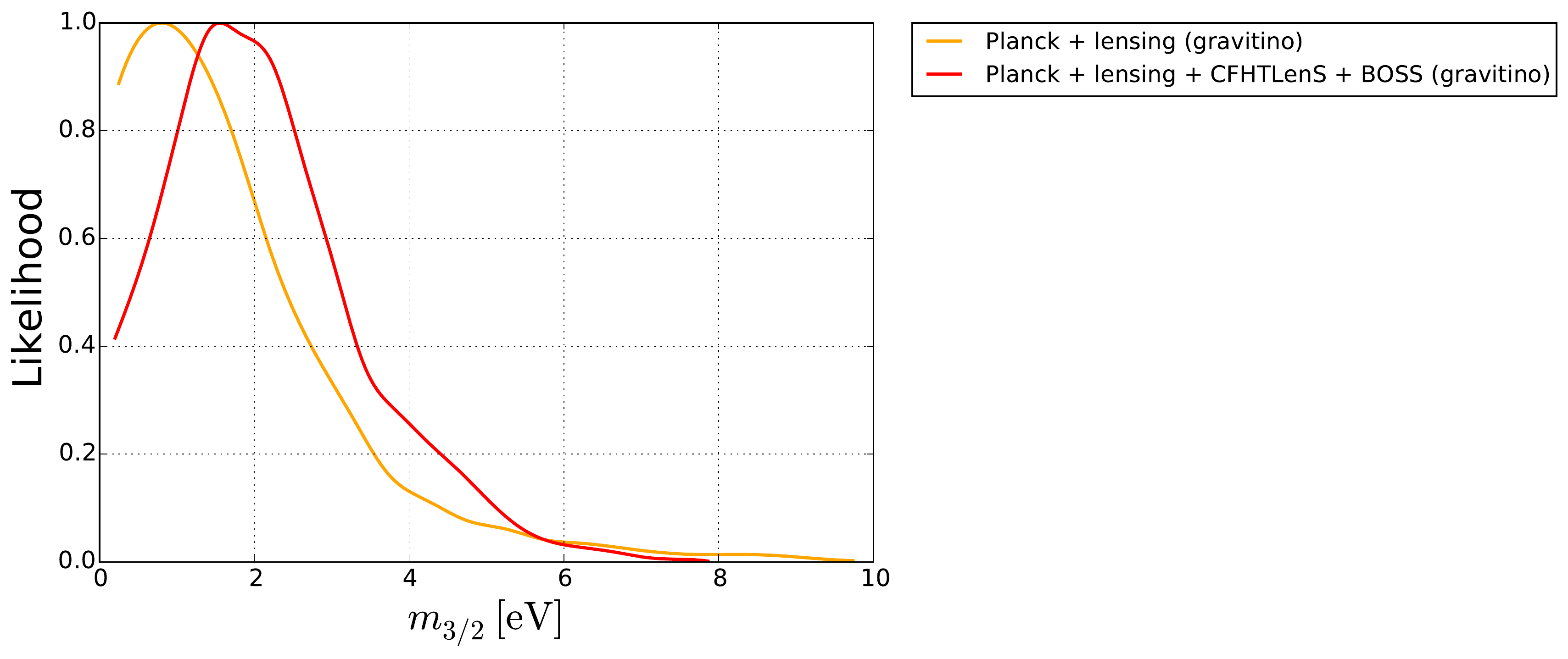}
\caption{
	Posterior distributions of the gravitino mass from two different data sets.
	In all figures we abbreviate ``Planck TT,TE,EE+lowP'' to ``Planck''.
}
\label{fig:grav_mass}
\end{figure}

\begin{figure}
\centering \includegraphics[clip, width=1.0\textwidth]{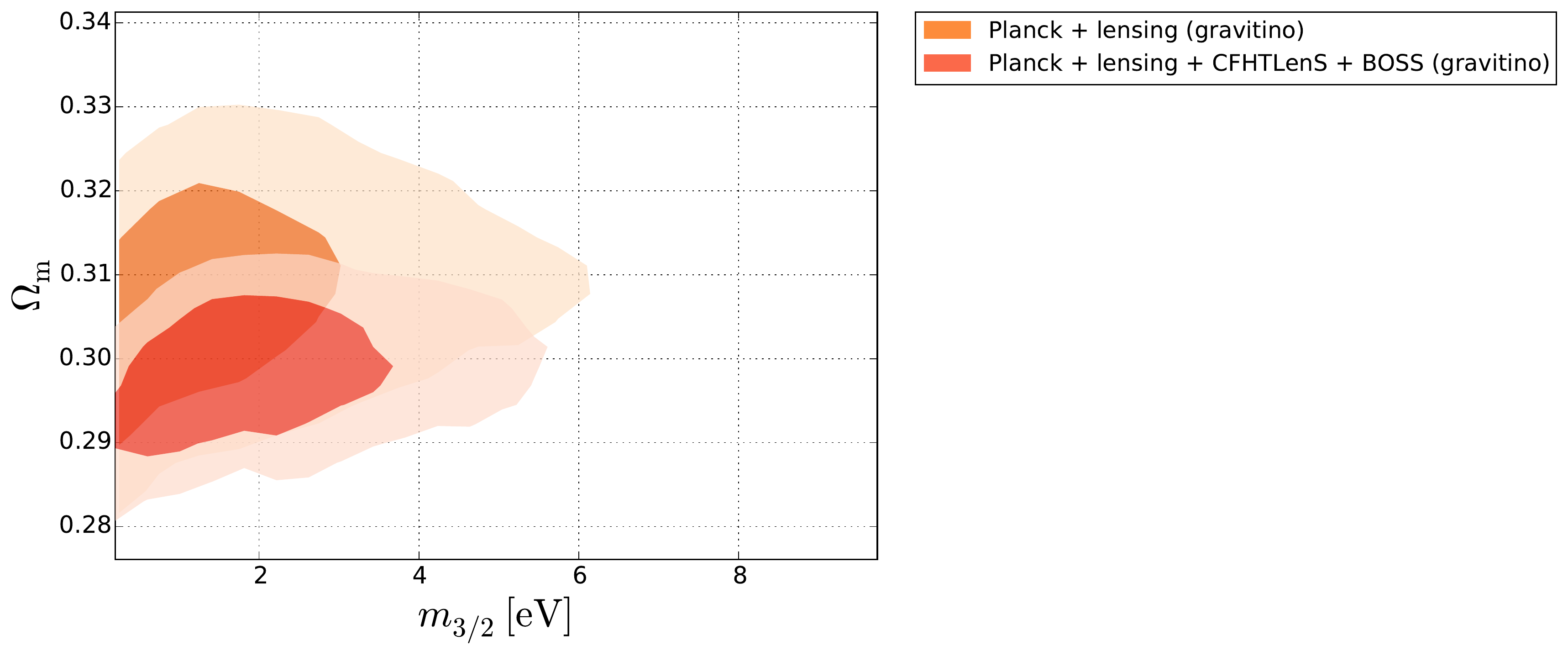}
\caption{
	Two-dimensional posterior distributions of the total matter density $\Omega_{\rm m}$
	and the gravitino mass  $m_{3/2}$
	from two different data sets.
}
\label{fig:Om_mgravitino}
\end{figure}

\begin{table}
  \centering
  \begin{tabular}{|c|c|} \hline
    Data set & $m_{3/2}\ [\mathrm{eV}]$ \\ \hline \hline
    Planck (TT,TE,EE+lowP)+lensing & $< 4.9$ (95\% C.L.) \\ \hline
    \shortstack{Planck (TT,TE,EE+lowP)+lensing\\+CFHTLenS+BOSS} & $2.1\pm1.2$ (68\% C.L.),\  $< 4.7$ (95\% C.L.) \\ \hline
  \end{tabular}
  \caption{Constraints on the mass of light gravitinos}
  \label{table:constraints}
\end{table}

Figure~\ref{fig:grav_mass} shows the posterior distributions of $m_{3/2}$
marginalized over all the other cosmological parameters,
and figure~\ref{fig:Om_mgravitino} shows the two-dimensional posterior distributions
of the total matter density and the gravitino mass.
Table~\ref{table:constraints} summarizes the constraints of $m_{3/2}$ with two different data sets.
Our upper bounds $m_{3/2}\lesssim 4.7$\,eV (95\% C.L.) are about three times more stringent
than the previous one derived from Ly-$\alpha$ forest observations, 
$m_{3/2} < 16~\mathrm{eV}$~\cite{Viel2005}.
When combined with the cosmic shear data, the width of the posterior distribution becomes twice smaller,
although the upper bounds are virtually unchanged.
Clearly adding cosmic shear observations is effective to constrain $m_{3/2}$,
although cosmic shear by itself alone cannot constrain $m_{3/2}$
very tightly (see appendix~\ref{sec:sigma8} and ref.~\cite{Kamada2014}).
We have examined the robustness of our results by removing the cosmic 
shear 2PCFs at small angular scales $\theta<10\ \mathrm{arcmin}$,
where the discrepancy between the simulation and HALOFIT results
is larger, while within 10\%.
We have confirmed that removing the small scale data leads to only small changes
and for the mass constraint, the difference is within $1 \sigma$.
The slight (only around 68\% C.L.) preference for nonzero $m_{3/2}$ in the ``all" data set 
arises from the tension in estimations of $\sigma_8$, which we shall discuss in appendix \ref{sec:sigma8}.

\section{Conclusions}
\label{sec:conc}
We have derived cosmological constraints on the gravitino mass $m_{3/2}=\mathcal O(1)$\,eV.
In the late-time universe, such light gravitinos can contribute to the total matter
density, and act as a warm component,
causing suppression of the matter density
fluctuations at $k=\mathcal{O} (0.01 \text{--} 0.1)\ \mathrm{Mpc}^{-1}$.
Interestingly, cosmological observations and collider experiments can
probe a SUSY breaking scale in a complemental way.

We have used two cosmological observations: 
the CMB lensing power spectrum from Planck,
and the two-point correlation function of cosmic shear from the CFHTLenS.
The combination enables us to measure the amplitude of the matter power spectrum
at a broad range of scales, which is essential to probe the light gravitino mass.
Combining the two data with primary CMB power spectra and galaxy clustering, 
we have obtained a stringent upper bound on the light gravitino mass
$m_{3/2}<4.7$\,eV (95\% C.L.).
Our constraint is considerably tighter than the previous constraint from 
Ly-$\alpha$ forest~\cite{Viel2005}. 

Measurements of cosmic shear will be improved significantly 
in the near future. High-quality data from, e.g.,
DES~\cite{2005astro.ph.10346T} \footnote{\rm{http://www.darkenergysurvey.org/}}
and Hyper Suprime-Cam~\cite{2012SPIE.8446E..0ZM} \footnote{\rm{http://www.naoj.org/Projects/HSC/index.html}}
will not only improve the present bound,
but also will allow us to explore a broader range of cosmological 
scenarios such as those with nonthermally produced gravitinos
with a low reheating temperature or ones diluted by a late entropy production.
We expect the future observations will provide us with 
knowledge of physics at high energy scales and in the early universe.

\acknowledgments
We thank Kiyotomo Ichiki for useful comments on the earlier version of the manuscript.
KO is supported by Advanced Leading Graduate Course for Photon Science.
MS is supported by Research Fellowships of the Japan Society for the Promotion of Science
(JSPS) for Young Scientists. 
KO, MS and NY acknowledge financial support from JST CREST.
TS is supported by IBS under the project code, IBS-R018-D1.
Numerical simulations were carried out on Cray XC30 at the Center for Computational Astrophysics,
National Astronomical Observatory of Japan.

\appendix
\section{Implications for $\sigma_8$ tension}
\label{sec:sigma8}

\begin{figure}
\centering \includegraphics[clip, width=1.0\textwidth]{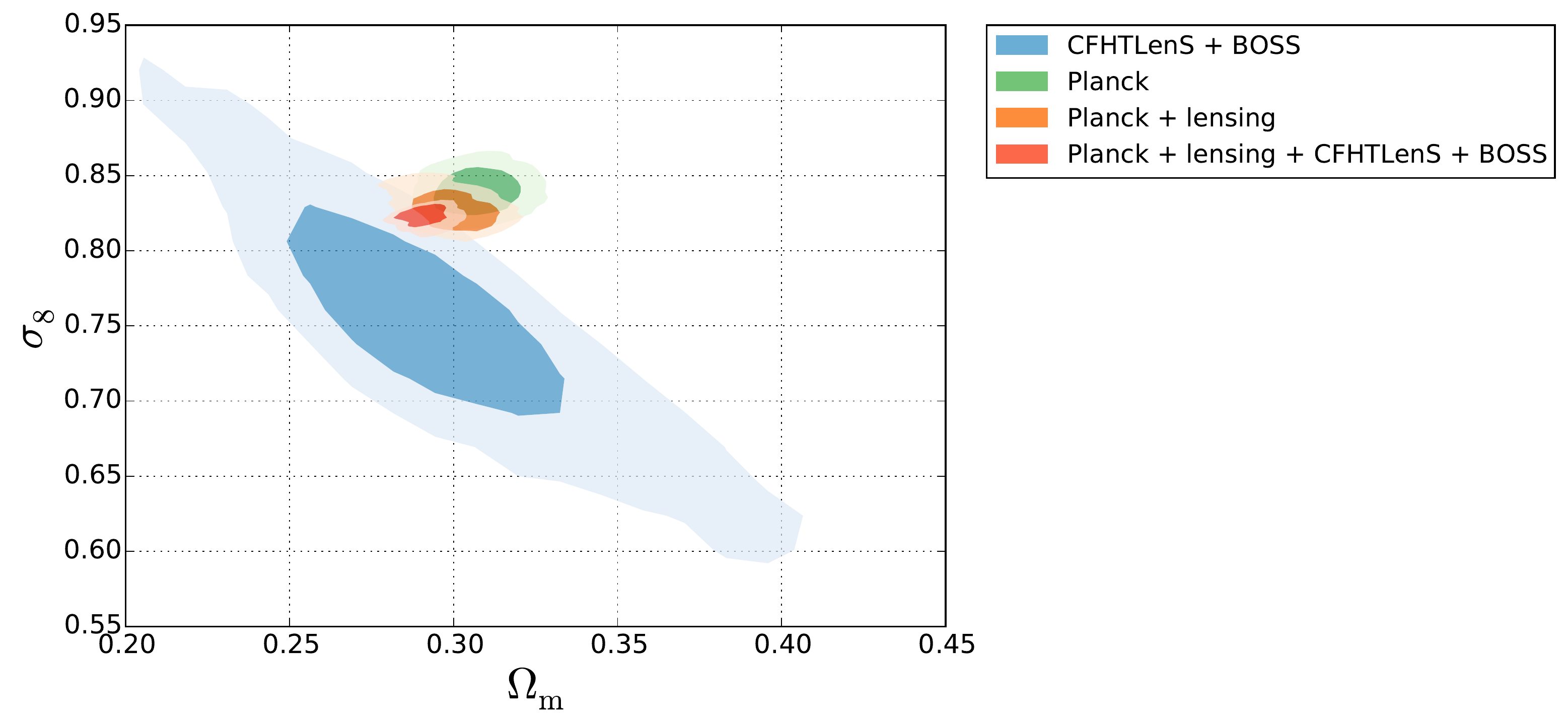}
\centering \includegraphics[clip, width=1.0\textwidth]{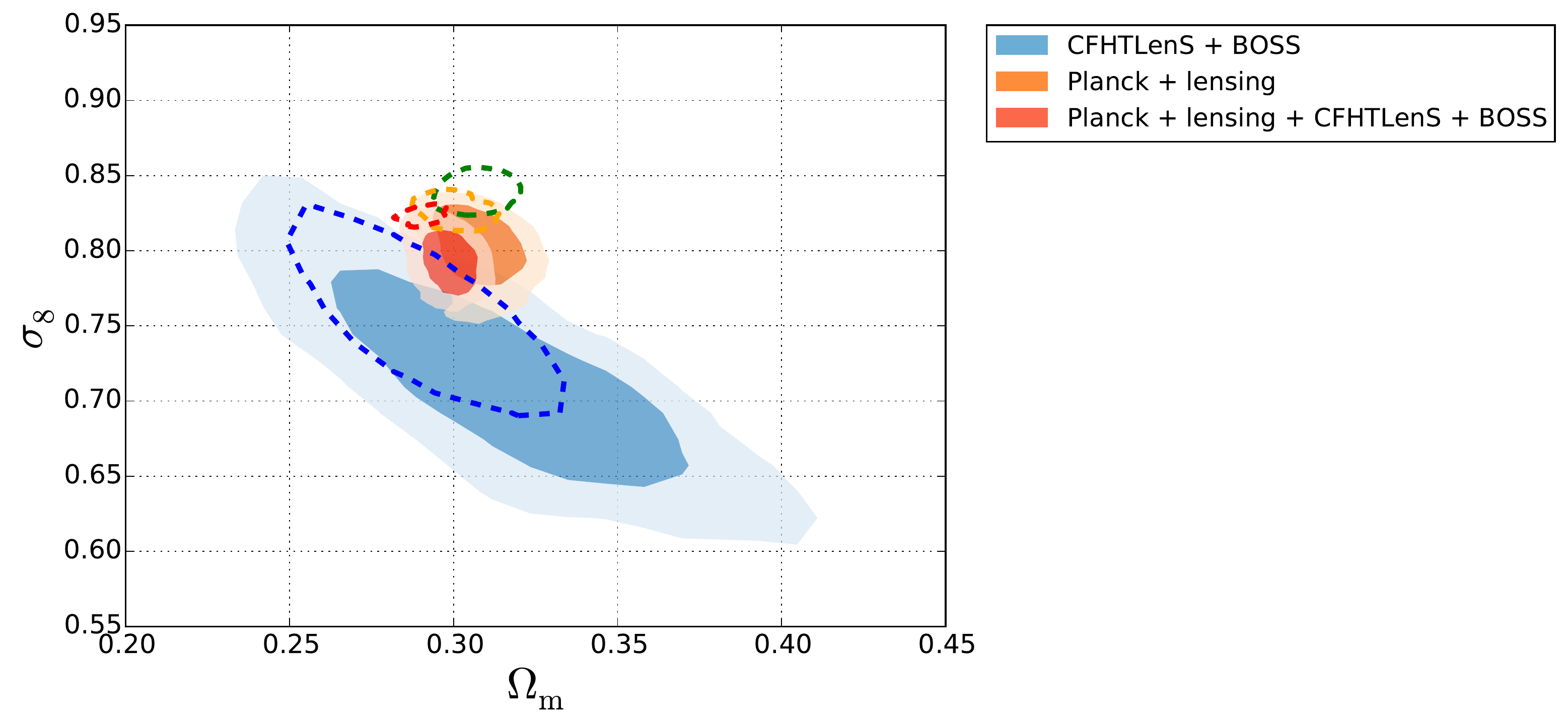}
\caption{
	Two-dimensional posterior distributions of the total matter density $\Omega_{\rm m}$
	and the amplitude of the fluctuation $\sigma_8$.
	The top and bottom panels show the cases without and with
	light gravitinos, respectively.
	For the purpose of comparison, in the bottom panel, the dashed lines 
	show the contours of distributions at 68\% C.L. results without light gravitinos.
}
\label{fig:area}
\end{figure}

\begin{figure}
\centering \includegraphics[clip, width=1.0\textwidth]{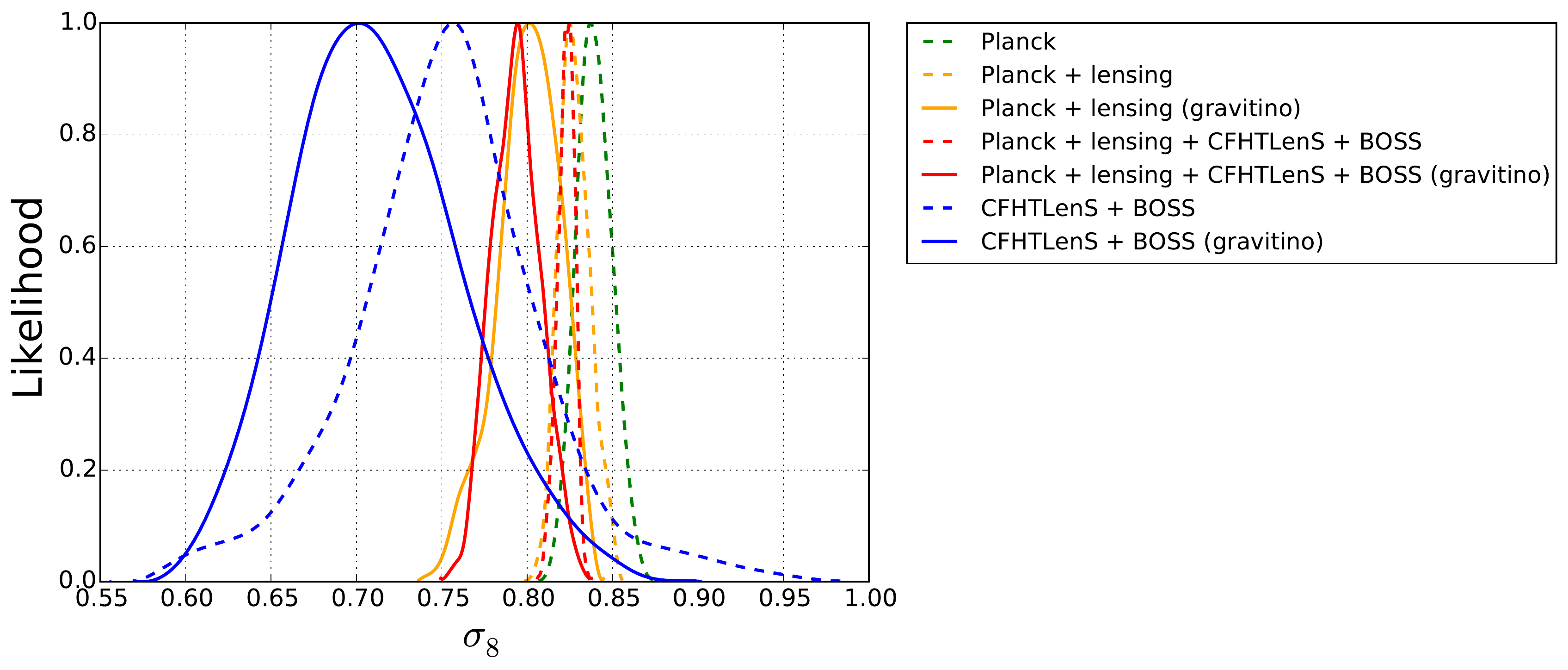}
\caption{
	Posterior distributions of $\sigma_8$ from different datasets. 
	Solid and dashed lines correspond to the cases with and without light gravitinos, respectively.
}
\label{fig:hist}
\end{figure}
In this appendix, we address the apparent tension in the estimate of 
$\sigma_8$.
We perform the MCMC analysis using only CHFTLenS and BOSS data.
Without the CMB, $m_{3/2}$ is almost unconstrained and can be
as large as $\mathcal{O}(10)$ eV. Because applying {\tt HALOFIT} to such models
can significantly compromise the parameter estimate,
we impose a top-hat prior on $m_{3/2}$ in [0,~10]~eV.
Since we use only the low-redshift large-scale structure probes, i.e., CFHTLenS and BOSS,
the baryon density and the optical depth 
are not constrained effectively.
We fix these parameters as the Planck 2015 TT,TE,EE+lowP values: 
$\Omega_\mathrm{b}h^2 = 0.02225,\ \tau_\mathrm{reio} = 0.0079$~\cite{PlanckParameters}.

Figure~\ref{fig:area} shows the contours of the two-dimensional posterior distribution of 
the total matter density $\Omega_{\rm m}=\Omega_{\rm b}+\Omega_{\rm cdm}+\Omega_{3/2}$
and the amplitude of the matter fluctuation $\sigma_8$ in models with and without light gravitinos.
The one-dimensional posterior distributions of $\sigma_8$ is also shown in figure~\ref{fig:hist}.
Light gravitinos suppress the matter power spectrum and effectively lower the fluctuation
amplitude as we have shown in sections \ref{sec:lss} and \ref{sec:probes}.
As a result, the distributions of $\sigma_8$ shift downward (figures~\ref{fig:area} and \ref{fig:hist}).
We have checked that removing small scale data ($\theta < 10\ \mathrm{arcmin}$) of 2PCFs
results in only minor changes
of the posterior distributions, and thus the main results remain robust if we use all the 2PCF data.
Although the tension between the two data sets with gravitinos,
CHFTLenS+BOSS (the blue solid line) and Planck+lensing (the yellow sold line),
is slightly mitigated, it is unlikely that invoking the
light gravitino helps reconcile the data at a sufficient level.

\bibliography{gravitino_library}

\providecommand{\href}[2]{#2}\begingroup\raggedright\begin{thebibliography}{10}

\bibitem{Martin:1997ns}
S.~P. {Martin}, {\it {a Supersymmetry Primer}},  {\em Perspectives On
  Supersymmetry.~Series: Advanced Series on Directions in High Energy Physics}
  {\bf 18} (July, 1998) 1--98, [\href{http://arxiv.org/abs/hep-ph/9709356}{{\tt
  hep-ph/9709356}}].

\bibitem{1982PhRvL..48.1303W}
S.~{Weinberg}, {\it {Cosmological constraints on the scale of supersymmetry
  breaking}},  {\em Physical Review Letters} {\bf 48} (May, 1982) 1303--1306.

\bibitem{1993PhLB..303..289M}
T.~{Moroi}, H.~{Murayama}, and M.~{Yamaguchi}, {\it {Cosmological constraints
  on the light stable gravitino}},  {\em Physics Letters B} {\bf 303} (Apr.,
  1993) 289--294.

\bibitem{1986PhLB..174...45F}
M.~{Fukugita} and T.~{Yanagida}, {\it {Barygenesis without grand unification}},
   {\em Physics Letters B} {\bf 174} (June, 1986) 45--47.

\bibitem{1998PhRvD..57.2089P}
E.~{Pierpaoli}, S.~{Borgani}, A.~{Masiero}, and M.~{Yamaguchi}, {\it {Formation
  of cosmic structures in a light gravitino-dominated universe}},  {\em Phys.
  Rev. D} {\bf 57} (Feb., 1998) 2089--2100,
  [\href{http://arxiv.org/abs/astro-ph/9709047}{{\tt astro-ph/9709047}}].

\bibitem{Viel2005}
M.~Viel, J.~Lesgourgues, M.~Haehnelt, S.~Matarrese, and A.~Riotto, {\it
  {Constraining warm dark matter candidates including sterile neutrinos and
  light gravitinos with WMAP and the Lyman-$\alpha$ forest}},  {\em Physical
  Review D} {\bf 71} (mar, 2005) 063534.

\bibitem{Ichikawa2009}
K.~Ichikawa, M.~Kawasaki, K.~Nakayama, T.~Sekiguchi, and T.~Takahashi, {\it
  {Constraining light gravitino mass from cosmic microwave background}},  {\em
  Journal of Cosmology and Astroparticle Physics} {\bf 2009} (aug, 2009)
  013--013.

\bibitem{Kamada2014}
A.~Kamada, M.~Shirasaki, and N.~Yoshida, {\it {Weighing the light gravitino
  mass with weak lensing surveys}},  {\em Journal of High Energy Physics} {\bf
  2014} (jun, 2014) 162.

\bibitem{2013JCAP...10..044H}
J.~{Hamann} and J.~{Hasenkamp}, {\it {A new life for sterile neutrinos:
  resolving inconsistencies using hot dark matter}},  {\em Journal of Cosmology
  and Astroparticle Physics} {\bf 10} (Oct., 2013) 44,
  [\href{http://arxiv.org/abs/1308.3255}{{\tt arXiv:1308.3255}}].

\bibitem{2014PhRvL.112e1303B}
R.~A. {Battye} and A.~{Moss}, {\it {Evidence for Massive Neutrinos from Cosmic
  Microwave Background and Lensing Observations}},  {\em Physical Review
  Letters} {\bf 112} (Feb., 2014) 051303,
  [\href{http://arxiv.org/abs/1308.5870}{{\tt arXiv:1308.5870}}].

\bibitem{2014MNRAS.444.3501B}
F.~{Beutler}, S.~{Saito}, J.~R. {Brownstein}, C.-H. {Chuang}, A.~J. {Cuesta},
  W.~J. {Percival}, A.~J. {Ross}, N.~P. {Ross}, D.~P. {Schneider},
  L.~{Samushia}, A.~G. {S{\'a}nchez}, H.-J. {Seo}, J.~L. {Tinker}, C.~{Wagner},
  and B.~A. {Weaver}, {\it {The clustering of galaxies in the SDSS-III Baryon
  Oscillation Spectroscopic Survey: signs of neutrino mass in current
  cosmological data sets}},  {\em Monthly Notices of the Royal Astronomical
  Society} {\bf 444} (Nov., 2014) 3501--3516,
  [\href{http://arxiv.org/abs/1403.4599}{{\tt arXiv:1403.4599}}].

\bibitem{Battye2015}
R.~A. Battye, T.~Charnock, and A.~Moss, {\it {Tension between the power
  spectrum of density perturbations measured on large and small scales}},  {\em
  Physical Review D} {\bf 91} (may, 2015) 103508,
  [\href{http://arxiv.org/abs/1409.2769}{{\tt arXiv:1409.2769}}].

\bibitem{MacCrann2015}
N.~MacCrann, J.~Zuntz, S.~Bridle, B.~Jain, and M.~R. Becker, {\it {Cosmic
  discordance: are Planck CMB and CFHTLenS weak lensing measurements out of
  tune?}},  {\em Monthly Notices of the Royal Astronomical Society} {\bf 451}
  (jun, 2015) 2877--2888.

\bibitem{Enqvist2015}
K.~Enqvist, S.~Nadathur, T.~Sekiguchi, and T.~Takahashi, {\it {Decaying dark
  matter and the tension in $\sigma_8$}},  {\em Journal of Cosmology and
  Astroparticle Physics} {\bf 2015} (sep, 2015) 067--067.

\bibitem{Osato2015}
K.~{Osato}, M.~{Shirasaki}, and N.~{Yoshida}, {\it {Impact of Baryonic
  Processes on Weak-lensing Cosmology: Power Spectrum, Nonlocal Statistics, and
  Parameter Bias}},  {\em The Astrophysical Journal} {\bf 806} (June, 2015)
  186, [\href{http://arxiv.org/abs/1501.02055}{{\tt arXiv:1501.02055}}].

\bibitem{Dine:1981gu}
M.~Dine and W.~Fischler, {\it {A Phenomenological Model of Particle Physics
  Based on Supersymmetry}},  {\em Phys. Lett.} {\bf B110} (1982) 227.

\bibitem{Nappi:1982hm}
C.~R. Nappi and B.~A. Ovrut, {\it {Supersymmetric Extension of the SU(3) x
  SU(2) x U(1) Model}},  {\em Phys. Lett.} {\bf B113} (1982) 175.

\bibitem{AlvarezGaume:1981wy}
L.~Alvarez-Gaume, M.~Claudson, and M.~B. Wise, {\it {Low-Energy
  Supersymmetry}},  {\em Nucl. Phys.} {\bf B207} (1982) 96.

\bibitem{Dine:1993yw}
M.~Dine and A.~E. Nelson, {\it {Dynamical supersymmetry breaking at
  low-energies}},  {\em Phys. Rev.} {\bf D48} (1993) 1277--1287,
  [\href{http://arxiv.org/abs/hep-ph/9303230}{{\tt hep-ph/9303230}}].

\bibitem{Dine:1994vc}
M.~Dine, A.~E. Nelson, and Y.~Shirman, {\it {Low-energy dynamical supersymmetry
  breaking simplified}},  {\em Phys. Rev.} {\bf D51} (1995) 1362--1370,
  [\href{http://arxiv.org/abs/hep-ph/9408384}{{\tt hep-ph/9408384}}].

\bibitem{Dine:1995ag}
M.~Dine, A.~E. Nelson, Y.~Nir, and Y.~Shirman, {\it {New tools for low-energy
  dynamical supersymmetry breaking}},  {\em Phys. Rev.} {\bf D53} (1996)
  2658--2669, [\href{http://arxiv.org/abs/hep-ph/9507378}{{\tt
  hep-ph/9507378}}].

\bibitem{Aad:2015zhl}
{\bf ATLAS, CMS} Collaboration, G.~Aad et~al., {\it {Combined Measurement of
  the Higgs Boson Mass in $pp$ Collisions at $\sqrt{s}=7$ and 8 TeV with the
  ATLAS and CMS Experiments}},  {\em Phys. Rev. Lett.} {\bf 114} (2015) 191803,
  [\href{http://arxiv.org/abs/1503.07589}{{\tt arXiv:1503.07589}}].

\bibitem{Chatrchyan:2012xdj}
{\bf CMS} Collaboration, S.~Chatrchyan et~al., {\it {Observation of a new boson
  at a mass of 125 GeV with the CMS experiment at the LHC}},  {\em Phys. Lett.}
  {\bf B716} (2012) 30--61, [\href{http://arxiv.org/abs/1207.7235}{{\tt
  arXiv:1207.7235}}].

\bibitem{Ajaib:2012vc}
M.~A. Ajaib, I.~Gogoladze, F.~Nasir, and Q.~Shafi, {\it {Revisiting mGMSB in
  Light of a 125 GeV Higgs}},  {\em Phys. Lett.} {\bf B713} (2012) 462--468,
  [\href{http://arxiv.org/abs/1204.2856}{{\tt arXiv:1204.2856}}].

\bibitem{Yanagida:2012ef}
T.~T. Yanagida, N.~Yokozaki, and K.~Yonekura, {\it {Higgs Boson Mass in Low
  Scale Gauge Mediation Models}},  {\em JHEP} {\bf 10} (2012) 017,
  [\href{http://arxiv.org/abs/1206.6589}{{\tt arXiv:1206.6589}}].

\bibitem{1996PhLB..389...37D}
S.~{Dimopoulos}, G.~F. {Giudice}, and A.~{Pomarol}, {\it {Dark matter in
  theories of gauge-mediated supersymmetry breaking}},  {\em Physics Letters B}
  {\bf 389} (Feb., 1996) 37--42,
  [\href{http://arxiv.org/abs/hep-ph/9607225}{{\tt hep-ph/9607225}}].

\bibitem{Aad:2014mra}
{\bf ATLAS} Collaboration, G.~Aad et~al., {\it {Search for supersymmetry in
  events with large missing transverse momentum, jets, and at least one tau
  lepton in 20 fb$^{-1}$ of $\sqrt{s}=$ 8 TeV proton-proton collision data with
  the ATLAS detector}},  {\em JHEP} {\bf 09} (2014) 103,
  [\href{http://arxiv.org/abs/1407.0603}{{\tt arXiv:1407.0603}}].

\bibitem{Matsumoto:2011fk}
S.~Matsumoto and T.~Moroi, {\it {Studying Very Light Gravitino at the ILC}},
  {\em Phys. Lett.} {\bf B701} (2011) 422--426,
  [\href{http://arxiv.org/abs/1104.3624}{{\tt arXiv:1104.3624}}].

\bibitem{Katayama:2013}
R.~Katayama, T.~Mori, K.~Fujii, S.~Matsumoto, T.~Suehara, T.~Tanabe, and
  S.~Yamashita, {\it {Full simulation study of very light gravitino at the
  ILC}},  {\em
  \href{http://flc.desy.de/lcnotes/noteslist/localfsExplorer_read?currentPath=/afs/desy.de/group/flc/lcnotes/LC-REP-2013-010.pdf}{LC-REP-2013-010}}
  (2013).

\bibitem{PlanckParameters}
{\bf Planck} Collaboration, P.~A.~R. Ade et~al., {\it {Planck 2015 results.
  XIII. Cosmological parameters}},  \href{http://arxiv.org/abs/1502.01589}{{\tt
  arXiv:1502.01589}}.

\bibitem{Lesgourgues2011}
J.~Lesgourgues, {\it {The Cosmic Linear Anisotropy Solving System (CLASS) I:
  Overview}},  {\em eprint arXiv:1104.2932} (2011).

\bibitem{Blas2011}
D.~Blas, J.~Lesgourgues, and T.~Tram, {\it {The Cosmic Linear Anisotropy
  Solving System (CLASS). Part II: Approximation schemes}},  {\em Journal of
  Cosmology and Astroparticle Physics} {\bf 2011} (jul, 2011) 034--034.

\bibitem{Lewis2006}
A.~Lewis and A.~Challinor, {\it {Weak gravitational lensing of the CMB}},  {\em
  Physics Reports} {\bf 429} (jun, 2006) 1--65.

\bibitem{Kaiser1992}
N.~Kaiser, {\it {Weak gravitational lensing of distant galaxies}},  {\em The
  Astrophysical Journal} {\bf 388} (apr, 1992) 272.

\bibitem{Hu:2000ee}
W.~Hu, {\it {Weak lensing of the CMB: A harmonic approach}},  {\em Phys. Rev.}
  {\bf D62} (2000) 043007, [\href{http://arxiv.org/abs/astro-ph/0001303}{{\tt
  astro-ph/0001303}}].

\bibitem{Bartelmann:1999yn}
M.~Bartelmann and P.~Schneider, {\it {Weak gravitational lensing}},  {\em
  Phys.Rept.} {\bf 340} (2001) 291--472,
  [\href{http://arxiv.org/abs/astro-ph/9912508}{{\tt astro-ph/9912508}}].

\bibitem{Munshi:2006fn}
D.~Munshi, P.~Valageas, L.~Van~Waerbeke, and A.~Heavens, {\it {Cosmology with
  Weak Lensing Surveys}},  {\em Phys.Rept.} {\bf 462} (2008) 67--121,
  [\href{http://arxiv.org/abs/astro-ph/0612667}{{\tt astro-ph/0612667}}].

\bibitem{Schneider:2002jd}
P.~Schneider, L.~van Waerbeke, M.~Kilbinger, and Y.~Mellier, {\it {Analysis of
  two-point statistics of cosmic shear: I. estimators and covariances}},  {\em
  Astron. Astrophys.} {\bf 396} (2002) 1--20,
  [\href{http://arxiv.org/abs/astro-ph/0206182}{{\tt astro-ph/0206182}}].

\bibitem{Takahashi2012}
R.~Takahashi, M.~Sato, T.~Nishimichi, A.~Taruya, and M.~Oguri, {\it {Revising
  the Halofit model for the nonlinear matter power spectrum}},  {\em The
  Astrophysical Journal} {\bf 761} (dec, 2012) 152.

\bibitem{Bird2012}
S.~Bird, M.~Viel, and M.~G. Haehnelt, {\it {Massive neutrinos and the
  non-linear matter power spectrum}},  {\em Monthly Notices of the Royal
  Astronomical Society} {\bf 420} (mar, 2012) 2551--2561,
  [\href{http://arxiv.org/abs/1109.4416}{{\tt arXiv:1109.4416}}].

\bibitem{PlanckLensing}
{\bf Planck} Collaboration, P.~A.~R. Ade et~al., {\it {Planck 2015 results. XV.
  Gravitational lensing}},  \href{http://arxiv.org/abs/1502.01591}{{\tt
  arXiv:1502.01591}}.

\bibitem{Kilbinger2013}
M.~Kilbinger, L.~Fu, C.~Heymans, F.~Simpson, J.~Benjamin, T.~Erben,
  J.~Harnois-D{\'{e}}raps, H.~Hoekstra, H.~Hildebrandt, T.~D. Kitching,
  Y.~Mellier, L.~Miller, L.~{Van Waerbeke}, K.~Benabed, C.~Bonnett, J.~Coupon,
  M.~J. Hudson, K.~Kuijken, B.~Rowe, T.~Schrabback, E.~Semboloni, S.~Vafaei,
  and M.~Velander, {\it {CFHTLenS: Combined probe cosmological model comparison
  using 2D weak gravitational lensing}},  {\em Monthly Notices of the Royal
  Astronomical Society} {\bf 430} (feb, 2013) 2200--2220,
  [\href{http://arxiv.org/abs/1212.3338}{{\tt arXiv:1212.3338}}].

\bibitem{Planck2013Parameters}
{\bf Planck} Collaboration, P.~A.~R. Ade et~al., {\it {Planck 2013 results.
  XVI. Cosmological parameters}},  {\em Astronomy {\&} Astrophysics} {\bf 571}
  (oct, 2014) A16.

\bibitem{Audren2013}
B.~Audren, J.~Lesgourgues, K.~Benabed, and S.~Prunet, {\it {Conservative
  constraints on early cosmology with MONTE PYTHON}},  {\em Journal of
  Cosmology and Astroparticle Physics} {\bf 2013} (feb, 2013) 001--001.

\bibitem{PlanckLikelihood}
{\bf Planck} Collaboration, P.~A.~R. Ade et~al., {\it {Planck 2015 results. XI.
  CMB power spectra, likelihoods, and robustness of parameters}},
  \href{http://arxiv.org/abs/1507.02704}{{\tt arXiv:1507.02704}}.

\bibitem{Okamoto:2003zw}
T.~Okamoto and W.~Hu, {\it {CMB lensing reconstruction on the full sky}},  {\em
  Phys. Rev.} {\bf D67} (2003) 083002,
  [\href{http://arxiv.org/abs/astro-ph/0301031}{{\tt astro-ph/0301031}}].

\bibitem{Erben:2012zw}
T.~Erben et~al., {\it {CFHTLenS: The Canada-France-Hawaii Telescope Lensing
  Survey - Imaging Data and Catalogue Products}},  {\em Mon. Not. Roy. Astron.
  Soc.} {\bf 433} (2013) 2545, [\href{http://arxiv.org/abs/1210.8156}{{\tt
  arXiv:1210.8156}}].

\bibitem{Heymans2012}
C.~Heymans, L.~{Van Waerbeke}, L.~Miller, T.~Erben, H.~Hildebrandt,
  H.~Hoekstra, T.~D. Kitching, Y.~Mellier, P.~Simon, C.~Bonnett, J.~Coupon,
  L.~Fu, J.~Harnois-D{\'{e}}raps, M.~J. Hudson, M.~Kilbinger, K.~Kuijken,
  B.~Rowe, T.~Schrabback, E.~Semboloni, E.~van Uitert, S.~Vafaei, and
  M.~Velander, {\it {CFHTLenS: the Canada--France--Hawaii Telescope Lensing
  Survey}},  {\em Monthly Notices of the Royal Astronomical Society} {\bf 427}
  (nov, 2012) 146--166.

\bibitem{Kirk:2011aw}
D.~Kirk, A.~Rassat, O.~Host, and S.~Bridle, {\it {The Cosmological Impact of
  Intrinsic Alignment Model Choice for Cosmic Shear}},  {\em Mon. Not. Roy.
  Astron. Soc.} {\bf 424} (2012) 1647,
  [\href{http://arxiv.org/abs/1112.4752}{{\tt arXiv:1112.4752}}].

\bibitem{York:2000gk}
{\bf SDSS} Collaboration, D.~G. York et~al., {\it {The Sloan Digital Sky
  Survey: Technical Summary}},  {\em Astron. J.} {\bf 120} (2000) 1579--1587,
  [\href{http://arxiv.org/abs/astro-ph/0006396}{{\tt astro-ph/0006396}}].

\bibitem{Ahn:2013gms}
{\bf SDSS-III} Collaboration, C.~P. Ahn et~al., {\it {The Tenth Data Release of
  the Sloan Digital Sky Survey: First Spectroscopic Data from the SDSS-III
  Apache Point Observatory Galactic Evolution Experiment}},  {\em Astrophys. J.
  Suppl.} {\bf 211} (2014) 17, [\href{http://arxiv.org/abs/1307.7735}{{\tt
  arXiv:1307.7735}}].

\bibitem{Alam:2015mbd}
{\bf SDSS-III} Collaboration, S.~Alam et~al., {\it {The Eleventh and Twelfth
  Data Releases of the Sloan Digital Sky Survey: Final Data from SDSS-III}},
  {\em Astrophys. J. Suppl.} {\bf 219} (2015), no.~1 12,
  [\href{http://arxiv.org/abs/1501.00963}{{\tt arXiv:1501.00963}}].

\bibitem{Anderson2014}
{\bf SDSS-III} Collaboration, L.~{Anderson} et~al., {\it {The clustering of
  galaxies in the SDSS-III Baryon Oscillation Spectroscopic Survey: baryon
  acoustic oscillations in the Data Releases 10 and 11 Galaxy samples}},  {\em
  Monthly Notices of the Royal Astronomical Society} {\bf 441} (June, 2014)
  24--62, [\href{http://arxiv.org/abs/1312.4877}{{\tt arXiv:1312.4877}}].

\bibitem{2005astro.ph.10346T}
{\bf The Dark Energy Survey} Collaboration, {The Dark Energy Survey
  Collaboration}, {\it {The Dark Energy Survey}},
  \href{http://arxiv.org/abs/astro-ph/0510346}{{\tt astro-ph/0510346}}.

\bibitem{2012SPIE.8446E..0ZM}
S.~{Miyazaki}, Y.~{Komiyama}, H.~{Nakaya}, Y.~{Kamata}, Y.~{Doi}, T.~{Hamana},
  H.~{Karoji}, H.~{Furusawa}, S.~{Kawanomoto}, T.~{Morokuma}, Y.~{Ishizuka},
  K.~{Nariai}, Y.~{Tanaka}, F.~{Uraguchi}, Y.~{Utsumi}, Y.~{Obuchi},
  Y.~{Okura}, M.~{Oguri}, T.~{Takata}, D.~{Tomono}, T.~{Kurakami},
  K.~{Namikawa}, T.~{Usuda}, H.~{Yamanoi}, T.~{Terai}, H.~{Uekiyo},
  Y.~{Yamada}, M.~{Koike}, H.~{Aihara}, Y.~{Fujimori}, S.~{Mineo},
  H.~{Miyatake}, N.~{Yasuda}, J.~{Nishizawa}, T.~{Saito}, M.~{Tanaka},
  T.~{Uchida}, N.~{Katayama}, S.-Y. {Wang}, H.-Y. {Chen}, R.~{Lupton},
  C.~{Loomis}, S.~{Bickerton}, P.~{Price}, J.~{Gunn}, H.~{Suzuki},
  Y.~{Miyazaki}, M.~{Muramatsu}, K.~{Yamamoto}, M.~{Endo}, Y.~{Ezaki},
  N.~{Itoh}, Y.~{Miwa}, H.~{Yokota}, T.~{Matsuda}, R.~{Ebinuma}, and
  K.~{Takeshi}, {\it {Hyper Suprime-Cam}},  in {\em Society of Photo-Optical
  Instrumentation Engineers (SPIE) Conference Series}, vol.~8446 of {\em
  Society of Photo-Optical Instrumentation Engineers (SPIE) Conference Series},
  p.~84460Z, Sept., 2012.

\end{thebibliography}\endgroup

\end{document}